\documentclass[preprint]{aastex}

\newcommand{\logg} {\log g}

\newcommand{\Te} {T_{\rm eff}}
\newcommand{\hm} {$\rm{H^-}$}
\newcommand{\mv} {$M_V$}
\newcommand{\htwo} {H$_2$}
\newcommand{\msun} {$M_\odot$}
\newcommand{\rsun} {$R_\odot$}
\newcommand{\lsun} {$L_\odot$}
\newcommand\gta{\lower 0.5ex\hbox{$\buildrel > \over \sim\ $}} 
\newcommand\lta{\lower 0.5ex\hbox{$\buildrel < \over \sim\ $}} 
\newcommand{\nhe} {N({\rm He})/N({\rm H})}
\newcommand{\nh} {N({\rm H})/N({\rm He})}
\newcommand{\nca} {N({\rm Ca})/N({\rm He})}
\newcommand{\WD} {WD 0346$+$246}
\shortauthors{Bergeron}
\shorttitle{\WD\ Revisited}
\begin{document}


\title{The Halo White Dwarf \WD\ Revisited}

\author{P. Bergeron}
\affil{D\'epartement de Physique, Universit\'e de Montr\'eal, C.P.~6128, 
Succ.~Centre-Ville, 
Montr\'eal, Qu\'ebec, Canada, H3C 3J7.}
\email{bergeron@astro.umontreal.ca}

\begin{abstract}

The extreme helium-rich atmospheric composition determined for the
halo white dwarf \WD\ is reexamined. This solution is shown to be
improbable from an astrophysical point of view when accretion of
hydrogen and metals from the interstellar medium is taken into
account. An alternate solution is proposed where hydrogen and helium
are present in the atmospheric regions in equal amounts. The best fit
at $\Te=3780$~K, $\logg=8.34$, and $\nhe=1.3$ is achieved by
including in the model calculations a bound-free opacity from the
Lyman edge associated with the so-called dissolved atomic levels of
the hydrogen atom, or pseudo-continuum opacity.

\end{abstract}


\section{Introduction}

The luminosity function of cool white dwarfs determined from proper
motion surveys \citep{ldm88,monet92} or colorimetric surveys
\citep{knox99} exhibits a sharp peak near $\log L/$\lsun$=-4.1$, followed by
a sudden drop towards lower luminosities. The paucity of white dwarfs
at low luminosities ($\log L/$\lsun$\sim-4.4$) has been interpreted as
a natural consequence of the finite age of the local Galactic
disk. These luminosity functions can be combined with theoretical
cooling sequences to derive an upper limit of 11 Gyr for the age of
the local disk
\citep[see the review of ][]{fon01}. Analyses of the coolest objects
in these surveys have revealed no evidence for white dwarfs cooler
than $\Te\sim4000$~K \citep{brl97,knox99,blr01}.

A handful of cooler and thus older white dwarfs, most likely belonging
to the thick disk or halo population, has recently been identified
\citep{hambly99,harris99,ibata00,harris01,opp01b}. This small sample
has been substantially increased by the exciting discovery by
\citet{opp01a} of 38 cool halo white dwarf candidates in the
SuperCOSMOS Sky Survey, although the halo nature of these objects has
recently been challenged by \citet{reid01}. The energy distributions
of the coolest white dwarfs in these samples are characterized by a
strong infrared flux deficiency and blue optical colors which have
been interpreted as the result of extremely strong
collision-induced absorptions by molecular hydrogen
\citep{hansen98,saumon99}.

Although the discovery of such faint and bluish white dwarfs confirms
the qualitative trends expected from model atmosphere calculations,
quantitative analyses of these objects have been less than successful,
in contrast to the detailed analyses of hotter ($\Te\ \gta 4000$~K)
white dwarfs in the Galactic disk \citep{brl97,blr01}.  For instance,
the energy distributions of LHS 3250 and its almost identical twin
SDSS 1337+00 show a maximum peak that is considerably broader and
bluer than predicted from pure hydrogen model atmospheres
\citep{harris99,harris01}. The discrepancy between the models and the
observations has been attributed to some missing opacity, most likely
originating from collision-induced absorptions by molecular hydrogen
due to collisions with {\it neutral helium}, although this interpretation
could not be confirmed quantitatively 
\citep{harris99,opp01b}. The temperature and composition of these two
white dwarfs thus remain unknown.

\citet{opp01b} provided a detailed analysis of the halo white dwarf 
\WD, as well as a preliminary analysis of F351$-$50,
discovered respectively by \citet{hambly97} and \citet{ibata00}. For
\WD, Oppenheimer et al.~obtained an excellent fit to the
observed optical $UBVRI$ and infrared $JHK$ photometry (see their
Fig.~8). Their solution suggests a cool ($\Te=3750$~K) and helium-rich
atmospheric composition, with an extremely small hydrogen abundance of
$\log \nh=-6.4$.

In this paper, we reexamine the solution proposed by Oppenheimer et
al. for \WD, and demonstrate from simple astrophysical considerations
that such a low hydrogen abundance is very unlikely. We first review
in \S~\ref{sec:cia} the importance of collision-induced opacities in
cool white dwarfs, and discuss the improvements to our model
atmosphere grid in \S~\ref{sec:mod}. The low hydrogen abundance
solution for \WD\ is then critically reevaluated in
\S~\ref{sec:hesol}. A modified treatment of the pseudo-continuum opacity from
the Lyman edge is introduced in
\S~\ref{sec:pseudo}, while our alternate solution using this previously
neglected source of opacity is discussed at length in
\S~\ref{sec:altsol}. Our conclusions follow in \S~\ref{sec:conc}.

\section{Collision-Induced Absorptions by \htwo\ in Cool White Dwarfs}\label{sec:cia}

The energy distributions of cool white dwarfs are characterized by a
strong infrared flux deficiency resulting from collision-induced
absorptions by molecular hydrogen (\htwo\ CIA). \citet{hansen98} has
recently demonstrated that the location of objects blueward of the
white dwarf cooling sequence in the \mv\ vs $(V-I)$ color-magnitude
diagram is actually the result of \htwo\ CIA which extends well into
the optical regions of the energy distribution of cool ($\Te\ \lta
3000$~K) and old ($\tau\ \gta 11$ Gyr) hydrogen atmosphere white
dwarfs \citep[see also][]{saumon99,blr01}. White dwarf stars thus
become increasingly redder as they cool down to $\Te\sim 4000$~K,
after which they evolve into bluer objects again. Recent papers
reporting the discovery of extremely cool white dwarfs have
perpetuated what has now become a common belief that the CIA opacity
is important only in this temperature range, however. It thus seems
appropriate here to first review our current knowledge of the
importance of \htwo\ CIA in cool white dwarf atmospheres.

In pure hydrogen atmospheres, the \htwo\ collision-induced opacity is
mainly due to collisions with other hydrogen molecules (\htwo-\htwo\
CIA). This absorption process becomes an important source of infrared
opacity in white dwarfs cooler than $\Te\sim5000$~K
\citep[see Fig.~5 of][]{bsw95}. At $\Te=4000$~K, for instance, more
than 50\% of the flux at 2 microns is absorbed by molecular hydrogen.
It is only below $\sim 4000$ K, however, that the
\htwo\ CIA opacity starts affecting the {\it optical regions} of the
energy distribution. Hence, optical color indices of extremely cool
white dwarfs, such as ($V-I$), become smaller (i.e., the objects
become bluer), as first pointed out by \citet{hansen98}. Color indices
in the infrared, however, become affected at much higher effective
temperatures, as shown in the ($B-V$, $V-K$) diagram of \citet[][their
Fig.~9]{brl97}.

Cool white dwarfs with mixed hydrogen and helium abundances are
characterized by even stronger infrared flux deficiencies. The
absorption by molecular hydrogen in this case is due to collisions
with both \htwo\ {\it and neutral helium} (\htwo-He CIA). In white
dwarf atmospheres where helium is the dominant constituent, \htwo-He
CIA may even represent the dominant source of opacity in the infrared
\citep[see Fig.~6 of][]{bsw95}. Even though 
the \htwo-\htwo\ and \htwo-He CIA absorption coefficients are
extremely similar \citep[see Fig.~7 of][]{jorgensen}, the atmospheric
pressures of mixed helium/hydrogen atmospheres are considerably larger
than those found in pure hydrogen atmospheres, and as such, the
collision-induced absorptions are correspondingly more important.
Consequently, the color-indices of mixed composition atmospheres
become affected by collision-induced absorptions at much higher
effective temperatures than those of pure hydrogen atmospheres, as
illustrated in the ($B-V$, $V-K$) two-color diagram shown in
\citet[][Fig.~9]{brl97}. Cool white dwarfs with mixed compositions thus easily 
stand out is such diagrams. LHS 1126 represents the first white dwarf
successfully interpreted in terms of \htwo-He CIA in a mixed
hydrogen/helium atmosphere \citep{ber94}; it is important to point out
in this context that the case of LHS 1126 is a definite confirmation
of the presence of \htwo-He CIA in a cool white dwarf, and not a {\it
possible} identification, as suggested by \citet{harris99}.

\section{Model Atmospheres}\label{sec:mod}

\citet{bsw95} calculated model atmospheres with mixed hydrogen and 
helium abundances of $\nhe=0.1$, 1, and 10. Theoretical colors
calculated from these models indicate that when the helium abundance
is increased, the photometric sequences further deviate from the pure
hydrogen sequence {\it as well as from the pure helium sequence}, as a
result of increased collision-induced absorptions. As first pointed
out by
\citet{bsw95}, the photometric sequences of models with even larger helium
abundances must at some point move back towards the pure helium
sequence. Cool model atmospheres with $\nh\ \lta10^{-3}$ could not be
calculated at that time, however, since the photospheric pressure that
characterizes such low hydrogen abundance atmospheres is so high that
the hydrogen atomic levels --- including that of the \hm\ ion --- are
strongly perturbed, and must be treated carefully within the
occupation probability formalism of \citet{hm88}, which had not been
included in these earlier models \citep[see \S~5.3 of][]{brl97}.

Details of our current model grid are discussed in
\citet{blr01} and references therein. In particular, the
Hummer-Mihalas formalism has now been included in detail following the
work of \citet{malo99}. This improvement now allows us to calculate
cool white dwarf models with arbitrarily low hydrogen
abundances. Also, the $\rm{He_2^+}$ ion has been included
self-consistently in the ionization equilibrium; as discussed by Malo
et al., the $\rm{He_2^+}$ ion may completely govern the electron
density in cool ($\Te\ \lta 8500$~K) helium-rich models.  For the
purpose of this study, we also make use of the recent {\it ab initio}
\htwo-He CIA calculations presented in \citet{jorgensen}. Earlier
white dwarf models used by \citet{brl97,blr01}, and those calculated
by \citet{opp01b}, all relied on approximate calculations that are
partially based on the work of Borysow \& Frommhold \citep[see][and
references therein]{lenzuni}. The differences between the new
calculations and the previous approximations are significant,
particularly at low temperatures \citep[see Figs.~1 and 6
of][]{jorgensen}.

Models with extremely low hydrogen abundances, $\nh\ \lta 10^{-7}$,
have also been presented in \citet{opp01b} and used to analyze the
cool white dwarf \WD, although no details of the physics used in these
models are provided.

The effects of helium abundance variations on the emergent fluxes of a
$\Te=3750$~K, $\logg=8.0$ white dwarf model --- i.e., the effective
temperature and surface gravity derived by \citet{opp01b} for \WD\ ---
are illustrated in Figure \ref{fg:f1}.  Although the model atmosphere
grid has been calculated with a resolution of 1 dex in $\log \nhe$,
only representative models are shown here.  As expected, the infrared
flux deficiency becomes more important as the helium abundance is
gradually increased, until it reaches a maximum at $\nhe=10^5$.
Simultaneously, the peak of the energy distribution moves towards
shorter wavelengths, and the slope of the optical regions becomes
steeper (the star appears bluer).  When the helium abundance is
further increased, \htwo-He CIA becomes only a small contribution to
the total opacity, the infrared depression weakens, and the peak of
the energy distribution rapidly shifts towards longer
wavelengths. Similar results were obtained by
\citet{opp01b} in two-color diagrams (see their Figs.~6 and 7),
although the convergence on the pure helium sequence that occurs at
$\nhe\ \gta 10^{13}$ in our models was not reached in their
analysis. Note that the most extreme helium abundance models discussed
here do not take into account the nonideal effects included in the
equation-of-state of the pure helium model grid calculated by
\citet{bsw95}, and as such, these models should be considered 
illustrative only.

\section{Reappraisal of the Low Hydrogen Abundance Solution}\label{sec:hesol}

We now attempt to fit the observed energy distribution of \WD\ using
the fitting technique described in \citet{blr01}. Briefly, the optical
$BVRI$ and infrared $JHK$ magnitudes taken from Table 1 of
\citet{opp01b} are first converted into average fluxes using equation
(1) of \citet{brl97}. These fluxes are then compared, using a
nonlinear least-squares method, with those obtained from the model
atmospheres, properly averaged over the filter bandpasses. Only $\Te$
and the solid angle $(R/D)^2$ are considered free parameters. The
distance $D$ is obtained from the trigonometric parallax measurement,
if available, and the stellar radius $R$ is converted into mass using
C/O-core cooling sequences described in \citet{blr01} with thin or
thick hydrogen layers, which are based on the calculations of
\citet{fon01}. For comparisons, however,  
we first adopt the temperature and surface gravity from
\citet{opp01b}.

Our best fit is presented in Figure \ref{fg:f2}. The top
panel shows the results in terms of $\log f_{\lambda}$, which is the
representation used by
\citet[][see their Fig.~8]{opp01b}, while the bottom panel is 
in terms of $f_{\nu}$, which is the way fits are displayed in
\citet{brl97,blr01}. Hence a good fit could be achieved with $\log \nhe=9.1$,
which is qualitatively similar to the best fit obtained in Figure 8 of
Oppenheimer et al. Our fit could even perhaps be improved by allowing
$\Te$ and $\logg$ to vary. We also note that the discrepancy
observed for the $I$ bandpass in the lower panel of Figure
\ref{fg:f2} is also present in the solution of Oppenheimer
et al., but this discrepancy becomes less apparent when the solution
is presented in terms of $\log f_{\lambda}$ (upper panel). Both
solutions differ quantitatively by nearly 3 orders of magnitude in
helium abundance, however. The origin of this difference is unknown,
but may stem from the differences in the input physics used in the two
model sets. The important aspect of both solutions nevertheless
remains that an {\it extremely high helium abundance is required to
fit the observed energy distribution of \WD}.

Such a low hydrogen content of only $\nh\sim10^{-9}$ is quite
unrealistic, however. Indeed, the white dwarf cooling age of \WD\
implied by our atmospheric parameter determination is 9.6 Gyr (using
the thick layer evolutionary models described above; thin layer models
yield 8.9 Gyr), during which the star has been traveling across the
Galaxy, accreting material from the interstellar medium, including
large amounts of hydrogen. Even though elements heavier than helium
will rapidly sink below the photosphere because of the efficient
gravitational chemical separation that exists in white dwarf
atmospheres, hydrogen will tend to float to the surface in the absence
of competing mechanisms. In other words, {\it the hydrogen abundance
can only increase with time}.  Since the superficial layers of cool
white dwarfs are strongly convective, however, hydrogen accreted from
the interstellar medium will be homogeneously mixed with the helium
convective zone. Below $\Te\sim 12,000$~K, the mass of this helium
convection zone is almost constant at $M_{\rm He-conv}\sim
10^{-6}~M_{\ast}$ \citep{tassoul90}. The hydrogen abundance determined
here for \WD\ thus implies a total mass of hydrogen accreted of
$1.2\times10^{-16}$ \msun, and an accretion rate from the interstellar
medium of $1.3\times 10^{-26}$
\msun\ yr$^{-1}$ (or $6.3\times 10^{-24}$ \msun\ yr$^{-1}$ if we use the
abundance value of Oppenheimer et al.). Such low accretion rates are
of course completely unrealistic and would require extremely efficient
screening mechanisms. For instance, the theoretical estimates of
\citet{wes79} suggest time-averaged accretion rates of $\sim10^{-17}$ \msun\ 
yr$^{-1}$. Similarly, the two-phase accretion model invoked by
\citet{dupuis93} to account for the presence of metals in cool
helium-rich white dwarfs assumes that accretion is most of the time
small, typically $10^{-20}$ \msun\ yr$^{-1}$, but may proceed at a
much higher rate of $5\times10^{-15}$ \msun\ yr$^{-1}$ during passages
through dense interstellar clouds.

The extreme purity of the atmosphere suggested for \WD\ also
represents a problem if the presence of metals is considered. Even
though elements heavier than helium will diffuse downward at the
bottom of the helium convection zone with time scales much shorter
than the evolutionary time scales
\citep[see][and references therein]{dupuis92}, magnesium, calcium, and
iron lines are nevertheless often observed in the spectra of cool,
helium-rich white dwarfs --- the so-called DZ stars. The only viable
explanation for the presence of these metals is provided by the
accretion-diffusion model in which a balance is achieved between the
gravitational settling of metals at the base of the helium convection
zone and the accretion of these elements at the surface of the star,
most of the time through the low-density interstellar medium, but
occasionally through denser ``clouds'' \citep[see, e.g.,][]{dupuis93}.

Model atmospheres have been calculated by including metals in the
equation-of-state only; metallic opacities, which are negligible at
the abundances and temperatures considered here, are not taken into
account. The main effect of the presence of heavy elements is to
provide free electrons, and to increase the contribution of the He$^-$
free-free opacity --- the dominant opacity source in pure helium
models. Here we include only one heavy element, calcium, the most
common metal observed in DZ stars. Calcium abundances determined in DZ
stars are typically $\nca=10^{-11}-10^{-10}$, although abundances as
high as $10^{-8}$ have also been observed in some objects
\citep{zeidler86}. Since calcium lines are not observed in the spectrum of \WD, 
we consider calcium abundances much lower than those determined in
typical DZ stars. The results of our calculations for white dwarf
models with $\Te=3750$~K, $\logg=8.0$, $\log \nhe=9.1$, and for
$\nca=0$, $10^{-14}$, $10^{-13}$, and $10^{-12}$ are displayed in
Figure
\ref{fg:f3}, together with the photometric observations of 
\WD\ shown for comparison. This experiment shows that even with a low calcium
abundance of only $\nca=10^{-14}$, the infrared fluxes are
significantly different from those obtained with the zero-metallicity model,
and different from the observed fluxes of \WD\ as well. Higher calcium
abundances --- and the inclusion of additional elements such as
magnesium or iron --- would probably make matters even worse.

We thus conclude that an almost pure helium atmospheric composition
for \WD, or any cool white dwarf in this temperature range for that
matter, represents an unlikely solution. In the following sections, we
propose an alternate solution for this object.

\section{Pseudo-Continuum Opacity from the Lyman Edge}\label{sec:pseudo}

\citet{brl97} showed that the photometric observations of DA 
(hydrogen-line) white dwarfs in the ($B-V$, $V-K$) two-color diagram
are well reproduced by pure hydrogen models down to $\Te\sim5000$~K,
below which the observed sequence seems to be better reproduced by the
pure helium models (see their Fig.~9). A closer examination of the
fits to the energy distributions of the cooler DA stars revealed that
a missing opacity source near the $B$ filter, resulting in an excess
of flux in this region, was responsible for this discrepancy. When
only wavelengths longward of $B$ were considered, such as in ($V-I$,
$V-K$) diagrams (see Fig.~13 of
\citealt{brl97} and Fig.~9 of \citealt{blr01}), the pure hydrogen models
followed perfectly the observed DA sequence.  For this reason, the
$B$ magnitude was excluded in the fits of the hydrogen-rich stars cooler
than $\Te\sim5500$~K.

\citet{brl97} proposed that this missing opacity source in the pure 
hydrogen models could be due to the bound-free opacity associated with
the so-called dissolved atomic levels of the hydrogen atom, or
pseudo-continuum opacity. When an electron makes a bound-bound
transition from a lower level to an upper level, there is a finite
probability that this upper level will be sufficiently perturbed by
surrounding particles (a {\it dissolved} level) such that the electron
is no longer bound to the nucleus, leading instead to a bound-free
transition. This pseudo-continuum opacity can be treated within the
Hummer-Mihalas formalism following the work of \citet{hm88} and
\citet{dappen}, which we briefly describe for completeness.

The contribution of level $i$ to the total monochromatic opacity is
given in LTE by 

$$\chi _i(\nu)=N_i \biggl[\sum_{j>i}{{w_j}\over {w_i}} \alpha _{ij}(\nu) 
+ D_i(\nu) \alpha_{i\kappa}(\nu) \biggr] (1-e^{-h\nu /kT}), \eqno (1)$$ 

\noindent where the first term corresponds to the bound-bound opacities weighted
by the occupation probability of each level of the transition. The
occupation probability $w_i$ of the atomic level $i$ corresponds to
the probability that an electron in this particular state is bound to
the atom relative to a similar atomic state in a non-interacting
environment. Correspondingly, $(1-w_i)$ is the probability that this
state belongs to the continuum due to interactions with neighboring
charged and neutral particles --- the so-called dissolved levels.

The second term in equation (1) corresponds to the pseudo-continuum
opacity, where $\alpha_{i\kappa}(\nu)$ is the photoionization
cross-section from level $i$, and $D_i(\nu)$ is the {\it dissolved fraction}
of levels given by the expression 

$$D_i(\nu)={{w_i-w_{n^\ast}}\over {w_i}}, \eqno (2)$$ 

\noindent where $n^\ast$ corresponds to the effective quantum number of the
atomic level reached after the absorption of a photon of energy
$h\nu$, i.e.,

$$n^\ast=\biggl( {1\over {n_i^2}}-{{h\nu}\over {\chi ^I_{\rm H}}} \biggr)
^{-{1\over 2}}. \eqno (3)$$ 

\noindent The atomic level $n^\ast$ is of course fictitious unless $n^\ast$ has
an integer value. An occupation probability $w_{n^\ast}$ is assigned
to that fictitious level by interpolating the value from the
occupation probability of the real atomic levels. When $n^\ast$ is
negative ($\nu >\chi^I_{\rm H}/hn_i^2$), we define $w_{n^\ast}\equiv
0$ (i.e., $D_i(\nu)=1$), and the transition becomes a true bound-free
transition. For positive values of $n^\ast$, $D_i(\nu)$ decreases with
increasing wavelength, and the product of this term with the
photoionization cross-section $\alpha_{i\kappa}(\nu)$ --- which is
simply extrapolated below the frequency threshold --- yields a
pseudo-continuum opacity that gradually fades away from the
unperturbed bound-free threshold. The pseudo-continuum opacity for the
Balmer jump in typical white dwarf atmosphere conditions is
illustrated in Figures 9 and 10 of \citet{ber91}.

A problem arises if one considers the pseudo-continuum opacity
originating from the Lyman edge. For instance, near 4000 \AA, the
dissolved fraction at the photosphere of a cool ($\Te\sim4500$~K),
hydrogen-rich white dwarf is typically $10^{-4}$. That is, electrons
in the ground level of hydrogen making a corresponding transition to
the fictitious level $n^\ast\sim1.14$ (setting $n_i=1$ and
$\lambda=4000$ \AA\ in equation [3]) have only a small 0.01\%
probability of going into the continuum, and to contribute to the
bound-free opacity. However, since the pseudo-continuum opacity also
depends on the population of hydrogen in the ground state, which is
typically $10^{20} {\rm cm}^{-3}$ under such conditions, it remains
the dominant opacity source, even in the infrared!

The problem is illustrated more quantitatively in Figure
\ref{fg:f4} where the contributions of the most relevant
sources of opacity in a cool hydrogen-rich atmosphere are shown. The
pseudo-continuum opacity from the $n=2$ level shows the expected
behavior: the bound-free cross section with an unperturbed threshold
at 3644 \AA\ is extended towards longer wavelength, and is gradually
attenuated by the dissolved fraction factor \citep[see also Fig.~2
of][]{dappen}. The behavior shown by the pseudo-continuum opacity from
the $n=1$ level (dotted line) is obviously unphysical, however. The
results would imply that this continuum opacity dominate all other
sources of opacity at all wavelengths! This is a familiar problem for
those who have attempted to implement the occupation probability
formalism of Hummer-Mihalas in white dwarfs cooler than
$\Te\sim17,000$~K, where most of the hydrogen lies in the ground
state. To overcome this problem, an arbitrary cutoff is usually
applied, or the pseudo-continuum opacity from the Lyman edge is
omitted altogether.

The exact reason why the Hummer-Mihalas formalism breaks down in this
particular situation is not known, and a detailed analysis of this
problem is outside the scope of this paper. We simply mention, as
discussed in
\citet{brl97}, that far away from the Lyman jump, the calculation of
the fraction of dissolved levels is probably unreliable. Indeed,
$D_i(\nu)$ is proportional to the difference between the occupation
probability of the ground level and that of the upper (fictitious)
dissolved level which, away from the edge, is located slightly above
the ground level. As such, $D(\nu)$ only {\it approaches} zero and its
exact value is meaningless, and so is the value of the corresponding
pseudo-continuum opacity.  The fact nevertheless remains that some form of
pseudo-continuum opacity from the Lyman edge must exist in
nature. This led \citet{brl97} to suggest that perhaps this neglected
opacity could be responsible for the observed discrepancy at the blue
end of the energy distribution of cool hydrogen-rich white
dwarfs. After all, the ratio of the threshold opacity for the $n=1$
level is more than 9 orders of magnitude larger than that of the $n=2$
level, and it is not too farfetched to expect some of this
pseudo-continuum opacity to affect the optical regions of the energy
distribution.

Despite the lack of an accurate theory for calculating the fraction of
dissolved levels far away from the Lyman edge, we may still make a
suitable approximation by including a ``damping'' function into the
pseudo-continuum opacity. Here we apply a damping function of
the form

$$D_i(\nu)^{\prime}=D_i(\nu) \exp\biggl(-{\Delta \lambda\over\Delta
\lambda_{\rm D}}\biggr)^a, \eqno (4)$$ 

\noindent where $a$ and $\Delta \lambda_{\rm D}$ are arbitrary damping factors
that remain to be determined. The results of the calculations with
$a=1$ and $\Delta \lambda_{\rm D}=350$ \AA\ are shown in Figure
\ref{fg:f4}. This particular choice remains of course
completely arbitrary but includes the two desired features,
i.e.~sufficient damping so that the overall energy distribution is
still dominated by H$^-$ in the optical and by \htwo\ CIA in the
infrared (not seen in Fig.~\ref{fg:f4}), and not too much
damping so that the pseudo-continuum opacity contributes to the total
opacity near the blue end of the optical region.

\section{Alternate Solution for \WD}\label{sec:altsol}

Energy distributions calculated with this modified pseudo-continuum
opacity are shown in Figure \ref{fg:f5} for two cool white
dwarfs taken from the analysis of \citet[][note that for this
comparison, the opacity has been included only in the calculation of
the emergent fluxes but not in the model structure itself]{blr01}. LP
380$-$5 (WD 1345$+$238) is a DA white dwarf cooler than the
temperature threshold at $\Te=5500$~K below which the observed $B$ magnitude of
hydrogen-rich white dwarfs in the studies of \citet{brl97,blr01} is
not well reproduced by the pure hydrogen models. The mismatch at $B$
for this object is also illustrated in Figure 24 of
\citet{brl97}. LHS 2522 (WD 1208$+$576) is another DA star with a
temperature slightly above this 5500~K threshold, and for which the
fit at $B$ is excellent. For LHS 2522, the inclusion of the
pseudo-continuum opacity from the Lyman edge has reduced the
theoretical fluxes for wavelengths shorter than the $B$ bandpass, and
the fit to the energy distribution is thus not affected by this
additional opacity. For LP 380$-$5, however, the flux is considerably
reduced near the $B$ bandpass, and models that include the
pseudo-continuum opacity reproduce the observed $B$ magnitude quite
well, as opposed to the model fluxes where this opacity is neglected.

Models were also calculated for mixed hydrogen and helium compositions
in the temperature range of \WD. The pseudo-continuum opacity is more
important in this object than in the other two DA white dwarfs
discussed above, not only because of its lower effective temperature,
but mainly because of the increased atmospheric pressure resulting
from the presence of helium. For this particular experiment, it was
thus necessary to include self-consistently the pseudo-continuum
opacity in the calculations of the model structures. In contrast with
our previous fit, we now consider $\Te$, $\logg$, and $\nhe$ free
parameters.

Our best fit for \WD\ is shown in the top panel of Figure
\ref{fg:f5}. Also shown for comparison is the optical and
infrared spectra from \citet[][note that the fluxes from the optical
spectrum have been multiplied here by a factor of 1.08 in order to
match with the observed photometry]{opp01b}. The results indicate that
a good fit to the energy distribution of \WD\ can be achieved with a
temperature of $\Te=3830$~K, close to the previous solution at
$\Te=3750$~K, but with a helium-to-hydrogen abundance ratio close to
unity. In particular, the fit at $B$ and $V$ is greatly improved by
the inclusion of the pseudo-continuum opacity, with respect to the fit
obtained when this opacity is neglected. Our fit can also be
contrasted with that shown in Figure 15 of \citet{fon01} for the same
object, based on slightly different photometric observations, and
where the $B$ and $V$ magnitudes were simply omitted from the fit
because of the suspected missing opacity in this region.

A more careful examination of the results shown in Figure
\ref{fg:f5} also reveals that even though the
inclusion of a pseudo-continuum opacity from the Lyman edge provides
on overall good fit to the observed energy distribution of \WD, our
approximate treatment could still be improved. If we take the observed
spectrum at face value, the pseudo-continuum opacity seems to extend
at much longer wavelengths, and to have the wrong frequency
dependency. As an illustrative example, by using different factors
$a=0.52$ and $\Delta \lambda_{\rm D}=36$ \AA\ in equation (4), and by
repeating the complete fitting procedure described above, we obtain
the fit shown in Figure \ref{fg:f6}. The atmospheric parameters
obtained with this modified damping function --- $\Te=3780$~K,
$\logg=8.34$, $\nhe=1.3$ --- are comparable to those given in Figure
\ref{fg:f5}, but the resulting fit is significantly better. In
particular, the slope and extent of the pseudo-continuum opacity are
reproduced much better. The pseudo-contiuum opacity profile at the
photosphere of the corresponding model atmosphere is shown in the
bottom panel of Figure \ref{fg:f6} together with the other most
relevant opacity sources.

We also note that the surface gravity inferred from our best fit
has been constrained from the measured trigonometric parallax
$\pi=36\pm5$ mas
\citep{hambly99}, which implies a radius of $R=0.0100$ \rsun\ and a mass
of $M=0.80$ \msun. In contrast, the former solution shown in Figure
\ref{fg:f2} as well as that of \citet{opp01b} both {\it assumed} 
$\logg=8.0$, which yields a radius of $R=0.01256$ \rsun\ and a
distance of $D=35.5$ pc, inconsistent with the measured distance of
$D=27.8$ pc. Using the evolutionary models discussed in \citet{blr01}
with C/O cores, helium envelopes of $q({\rm He})=10^{-2}$, and
``thick'' outermost hydrogen layers of $q({\rm H})=10^{-4}$, we derive
an age for \WD\ of $\tau=11.0$ Gyr; ``thin'' layer models with of
$q({\rm H})=10^{-10}$ yield $\tau=8.7$ Gyr.

As discussed in \S~4, the extreme helium-rich composition proposed by
\citet{opp01b} for \WD\ implies accretion rates from the 
interstellar medium that are unrealistically small. Adopting instead
$\nhe=1$ as our best estimate, we derive a total hydrogen mass of
$2\times10^{-7}$ \msun. If \WD\ has evolved from an almost pure helium
atmosphere --- i.e.~a DB star, the total amount of hydrogen accreted
corresponds to a time-averaged accretion rate of $\sim2\times10^{-17}$
\msun\ yr$^{-1}$, in good agreement with the theoretical estimates of
\citet{wes79} and \citet{dupuis93}. Alternatively, \WD\ may have 
evolved from a thin hydrogen layer DA star that has been convectively
mixed below $\Te\sim12,000$~K. According to Figure 40 of
\citet{brl97}, the minimum helium-to-hydrogen abundance ratio that can
be achieved for a 0.8 \msun\ white dwarf is when the hydrogen
atmosphere with a mass of $M({\rm H})\sim2.5\times10^{-8}$ \msun\ is
convectively mixed at $\Te\sim5000$~K. This yields an abundance of
$\nhe\sim8$. Hence, to arrive at an abundance of $\nhe=1$ by the time
the star has cooled down to $\Te=3800$~K, it must accrete an extra
$\sim1.8\times10^{-7}$
\msun\ of hydrogen over a period of $\sim1.6$ Gyr (assuming here thin 
hydrogen models), which corresponds to an accretion rate of $\sim10^{-17}$
\msun\ yr$^{-1}$. Hence it is not possible from these arguments alone
to determine whether \WD\ has evolved from a thin hydrogen-layer DA white
dwarf, or from a hot DB star.

We also discussed the fact that the extreme helium-rich composition
solution for \WD\ was unlikely because of the effects the presence of
even small amounts of elements heavier than helium have on the
predicted fluxes (see Fig. \ref{fg:f3}). Our solutions presented
in Figures \ref{fg:f5} or \ref{fg:f6} are not sensitive
to the presence of these additional elements since hydrogen remains
the most important contributor of free electrons. Additional
calculations not shown here indicate that the emergent fluxes are not
affected even when the calcium abundance is set to a value as high as
$\nca=10^{-8}$. Our solution is thus robust against the presence of
such additional elements.

\section{Conclusions}\label{sec:conc}

In this paper we have reexamined the atmospheric parameters of the
halo white dwarf \WD. A solution at $\Te\sim3750$~K with an extremely
low hydrogen abundance has been found, $\nh=10^{-9}$, in agreement
with the conclusions of \citet{opp01b}. We have showed, however, that
such a low hydrogen content is very unlikely due to accretion from the
interstellar medium over a cooling age of roughly 10 Gyr. Furthermore,
this solution was shown to be extremely volatile when even minute
amounts of heavy elements were included in the atmosphere.
We have proposed instead a solution where hydrogen and helium are
present in the atmosphere of \WD\ in nearly equal amounts. A good fit
has been achieved by including a pseudo-continuum opacity originating
from the dissolved levels of hydrogen. This opacity has been
parameterized with an {\it ad hoc} damping function, however, and it
is clear that a better physical description must be sought before more
quantitative results can be achieved.

The presence of this missing opacity source had already been
foreseen in the study of cool hydrogen-rich white dwarfs by
\citet{brl97,blr01} who refrained from interpreting
two-color diagrams where one of the filters may be affected by this
missing opacity, no matter what its origin might be --- ($B-V$, $V-K$)
diagrams for instance. These studies have also shown that for the
coolest ($\Te\sim4000$~K) hydrogen-rich stars, even the $V$ filter may
be affected. The results presented in this paper also support this
conclusion (see Fig.~\ref{fg:f5}). It thus appears extremely
dangerous to interpret these newly discovered ultracool white dwarfs
in two-color or color magnitude diagrams which involve ``blue''
magnitudes \citep[see, e.g.,][]{harris99,harris01,opp01a,opp01b} until
this problem with the missing opacity has been properly dealt with,
and included in the theoretical models and colors.

It now remains to be seen whether this new opacity source, even when
included with the approximate treatment introduced in this paper, can
help resolve the mystery surrounding several ultracool white dwarfs
whose energy distributions have yet failed to be successfully
explained in terms of hydrogen or mixed hydrogen/helium compositions.

\acknowledgements We are grateful to S.~T.~Hodgkin for providing us with 
the spectrum of \WD, and to G.~Fontaine and F.~Wesemael for a careful
reading of the manuscript. This work was supported in part by the
NSERC Canada and by the Fund FCAR (Qu\'ebec).

\clearpage
\figcaption[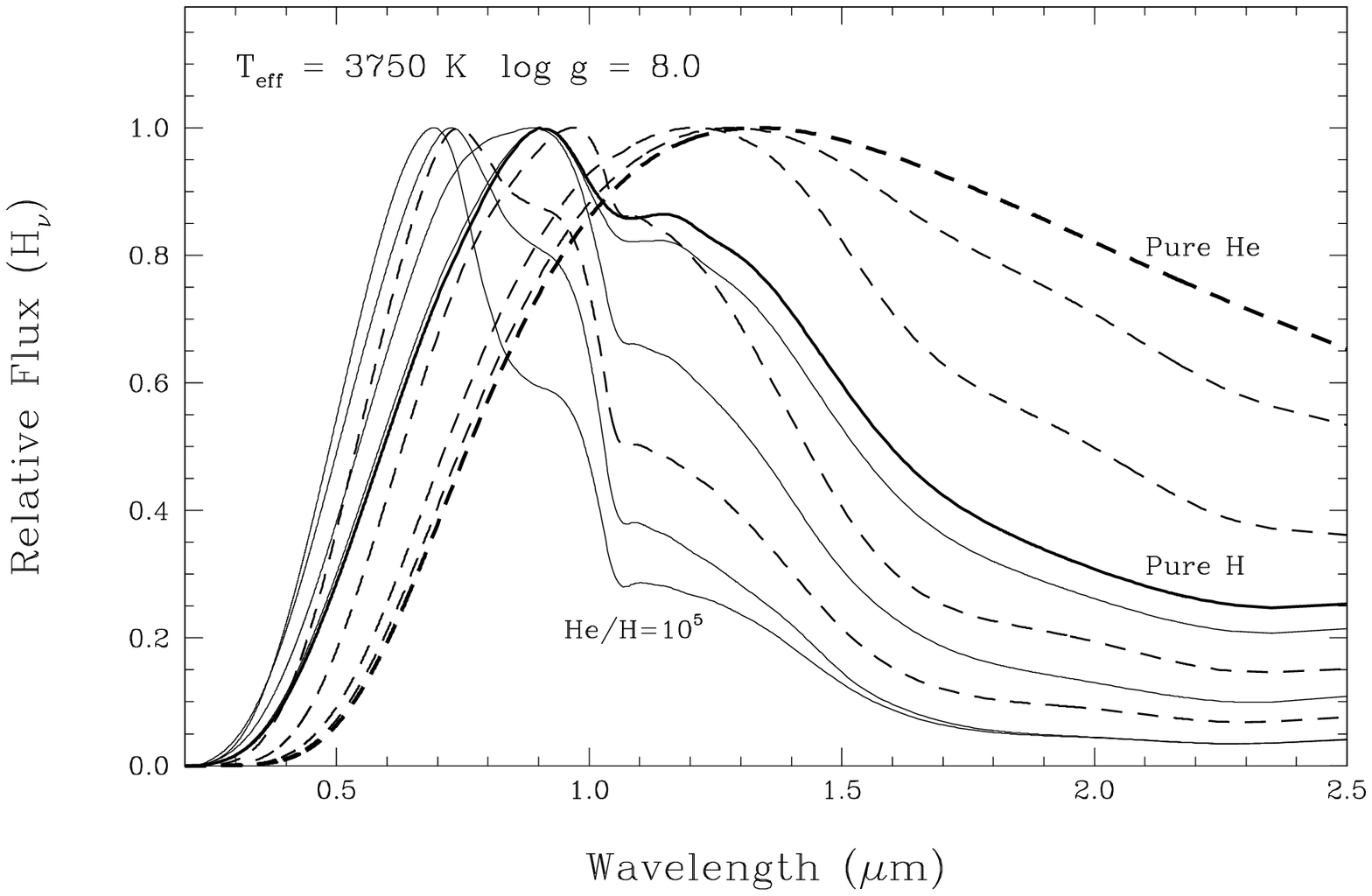] {Energy distribution 
of white dwarf model atmospheres with $\Te=3750$~K, $\logg=8.0$, and
various helium-to-hydrogen abundance ratios. All distributions are
normalized to unity where the Eddington flux ($H_{\nu}$)
is maximum. The thick solid and dashed lines represent respectively
the pure hydrogen and pure helium atmospheric compositions. The thin
solid lines correspond to (starting from the pure hydrogen model)
$\nhe=0.1$, 1, 10, and $10^5$, while the thin dashed lines correspond
to (starting from the $\nhe=10^5$ model) $\nhe=10^8$, $10^{10}$,
$10^{12}$, and $10^{13}$.\label{fg:f1}}

\figcaption[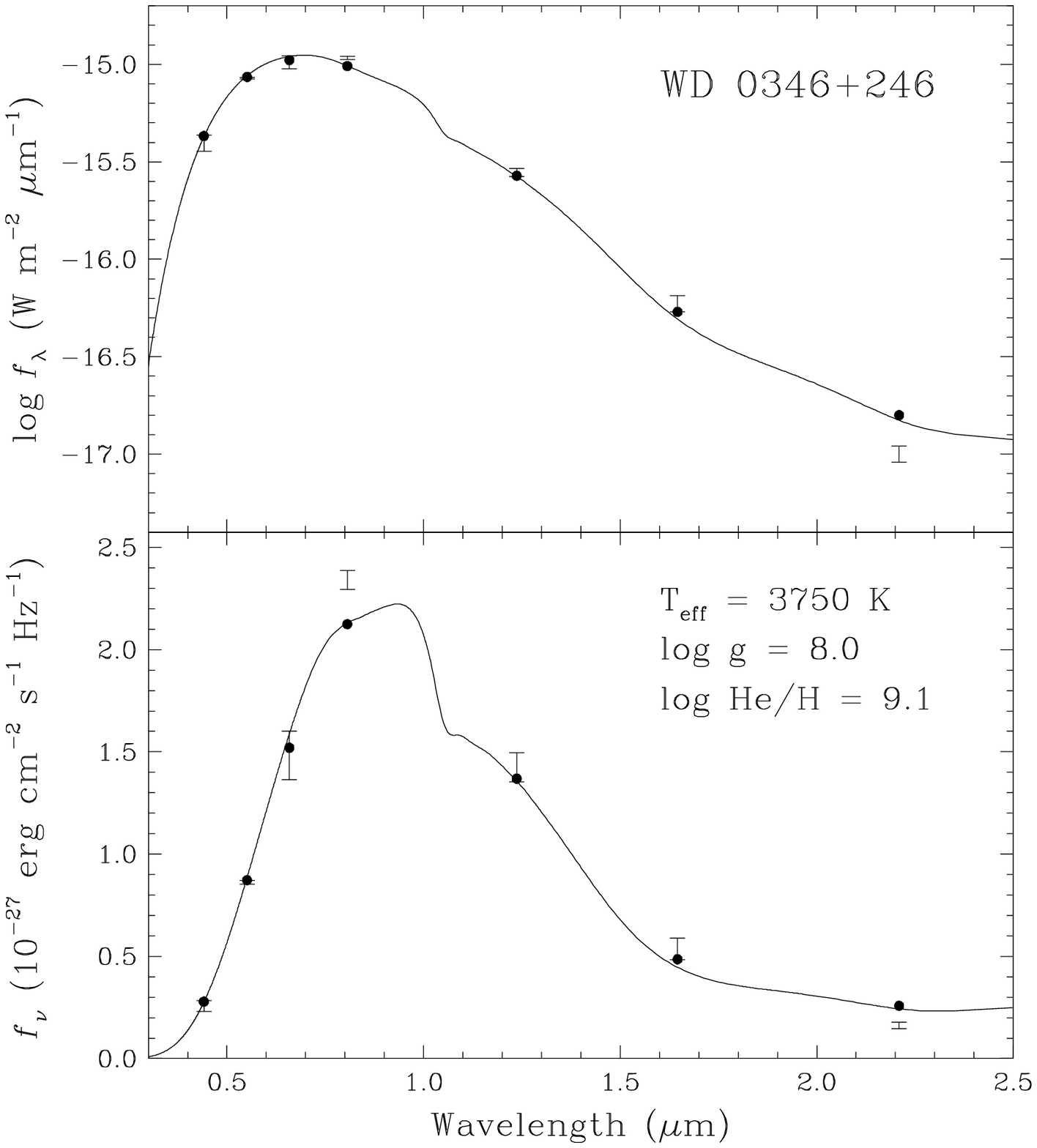] {Our best fit to the energy distribution 
of \WD\ in terms of $\log f_{\lambda}$ (top panel) and $f_{\nu}$ (bottom
panel); $\Te$ and $\logg$ are taken from
\citet{opp01b}. The $BVRI$ and $JHK$ photometric observations are
represented by error bars. The solid lines represent the model
monochromatic fluxes, while the filled circles correspond to the model
fluxes averaged over the filter bandpasses.\label{fg:f2}}

\figcaption[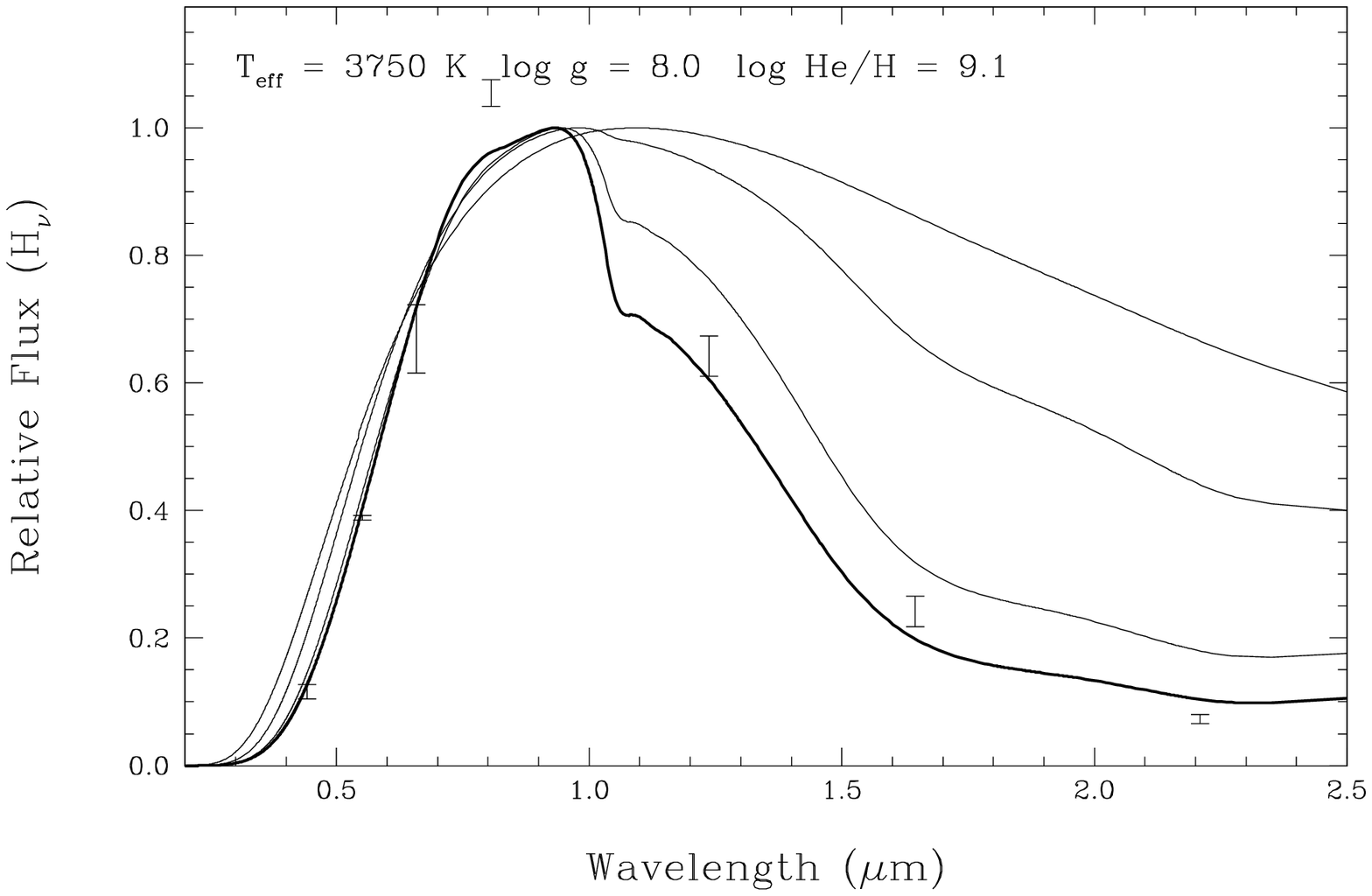] {Energy distribution 
of white dwarf model atmospheres with $\Te=3750$~K, $\logg=8.0$, $\log
\nhe=9.1$, and for various calcium-to-helium abundance ratios. All
distributions are normalized to unity where the Eddington flux
($H_{\nu}$) is maximum. The various models correspond to (starting
from the thick line) $\nca=0$, $10^{-14}$, $10^{-13}$, and
$10^{-12}$. The observed photometry of \WD, normalized with
the scaling factor determined from the zero-metal solution (see
Fig.~\ref{fg:f2}), is represented by error
bars.\label{fg:f3}}

\figcaption[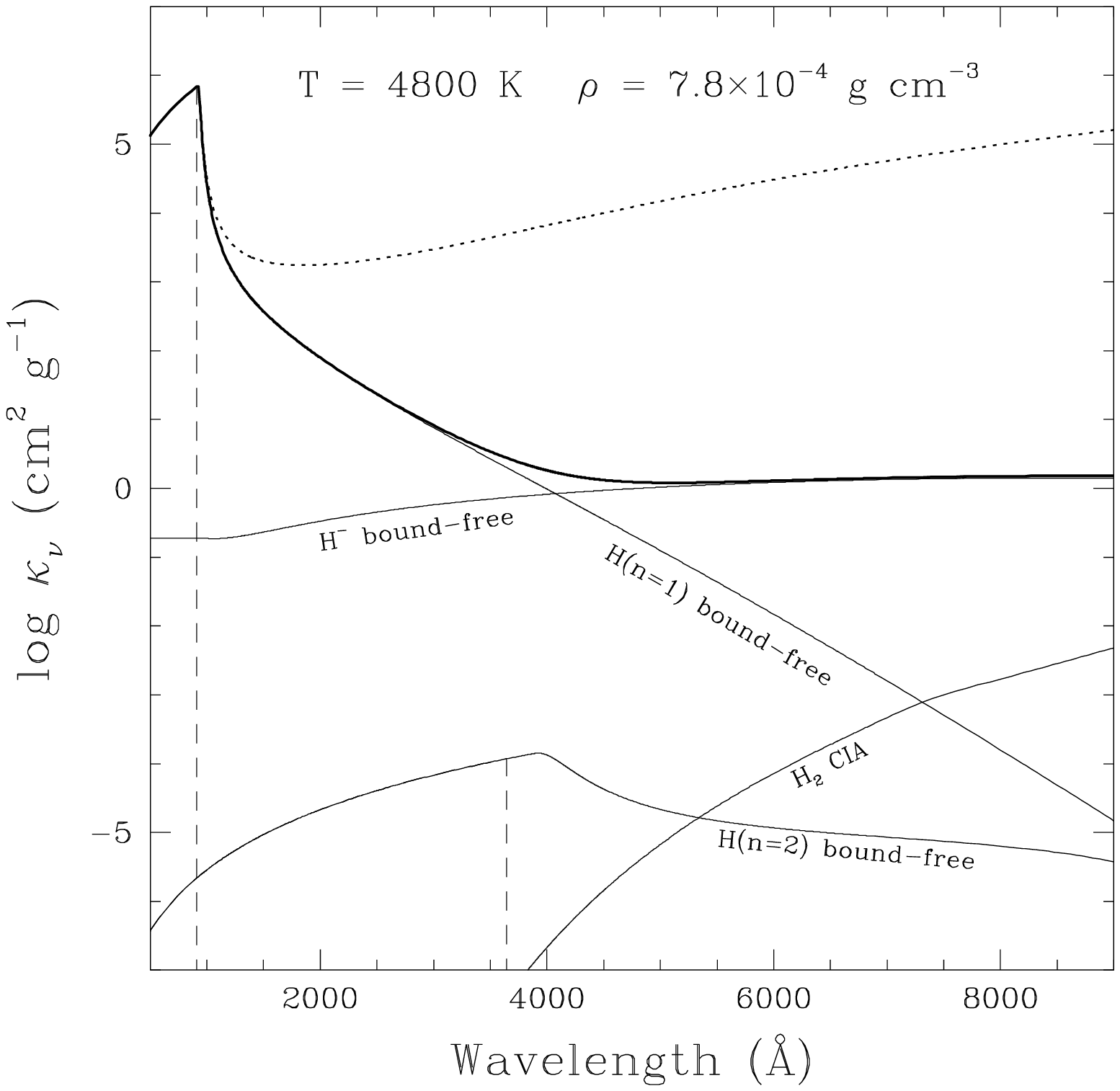] {Most relevant sources of opacity at the 
photosphere ($\tau_{\rm R}=1$) of a pure hydrogen model atmosphere at
$\Te=4600$~K and $\logg=7.75$ (the atmospheric parameters of LP
380$-$5 shown in Fig.~\ref{fg:f5}). The local temperature and
density are indicated in the figure, and each opacity source is
identified as well. The thick solid line represents the total
opacity. The bound-free opacities from the $n=1$ and $n=2$ levels
include the pseudo-continuum opacity from the dissolved levels, while
the contributions from the unperturbed bound-free absorptions alone
are represented by the dashed lines. The pseudo-continuum opacity from
the $n=1$ level calculated explicitly by equations (1) to (3) is shown
as a dotted line, and that calculated with the damping function given by
equation (4) with $a=1.0$ and $\Delta \lambda_{\rm D}=350$ \AA\ is
shown by the solid line.\label{fg:f4}}

\figcaption[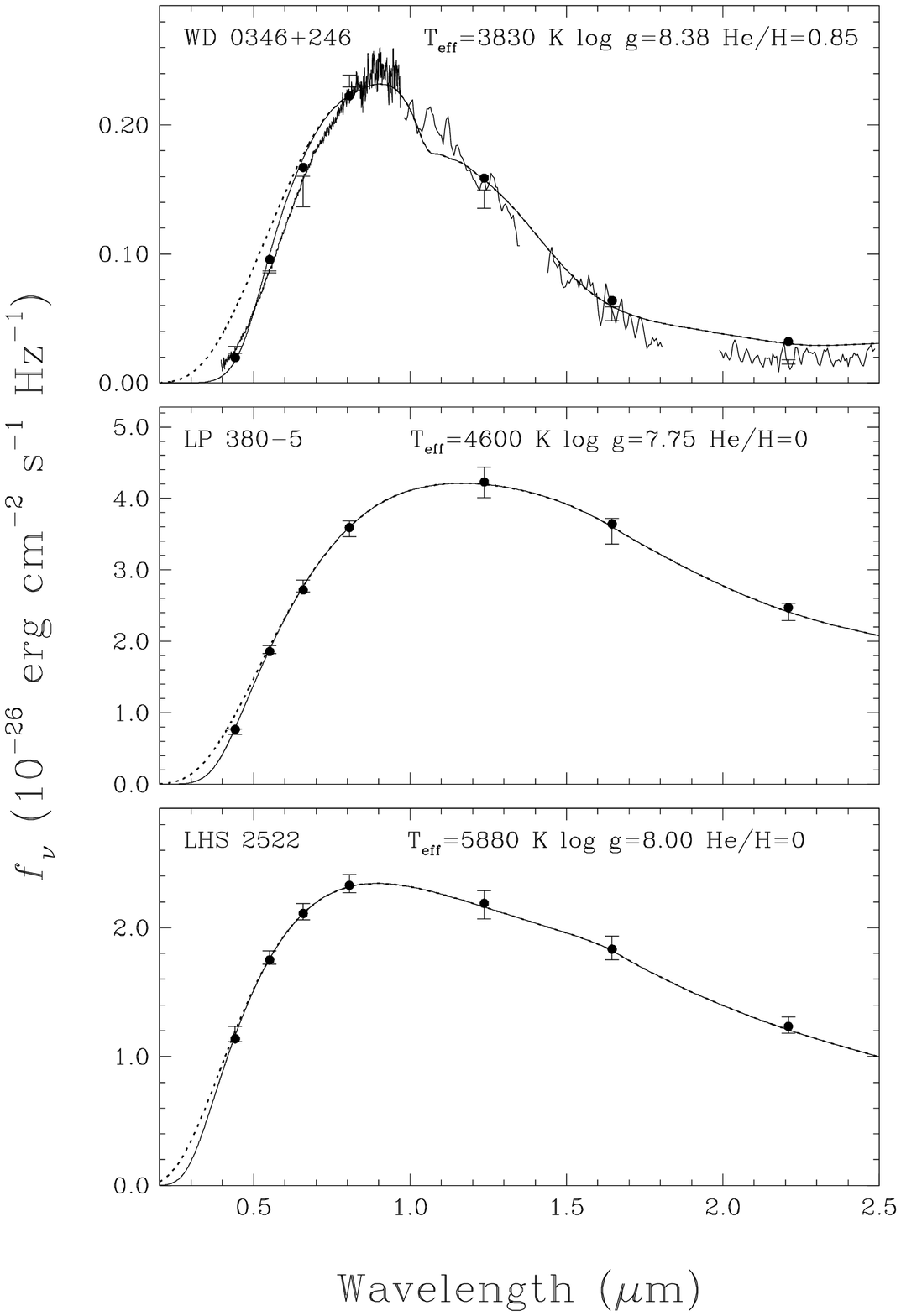] {Our best fit to the energy distribution
of \WD, as well as fits to LHS 2522 and LP 380$-$5 taken from
\citet{blr01}. The $BVRI$ and $JHK$ photometric observations are
represented by error bars. The solid and dotted lines represent
respectively the model fluxes with and without the pseudo-continuum
opacity from the Lyman edge, while the filled circles correspond to
the model fluxes --- with the pseudo-continuum opacity included ---
averaged over the filter bandpasses. The upper panel also shows for
comparison the optical and infrared spectra from
\citet{opp01b}.\label{fg:f5}}

\figcaption[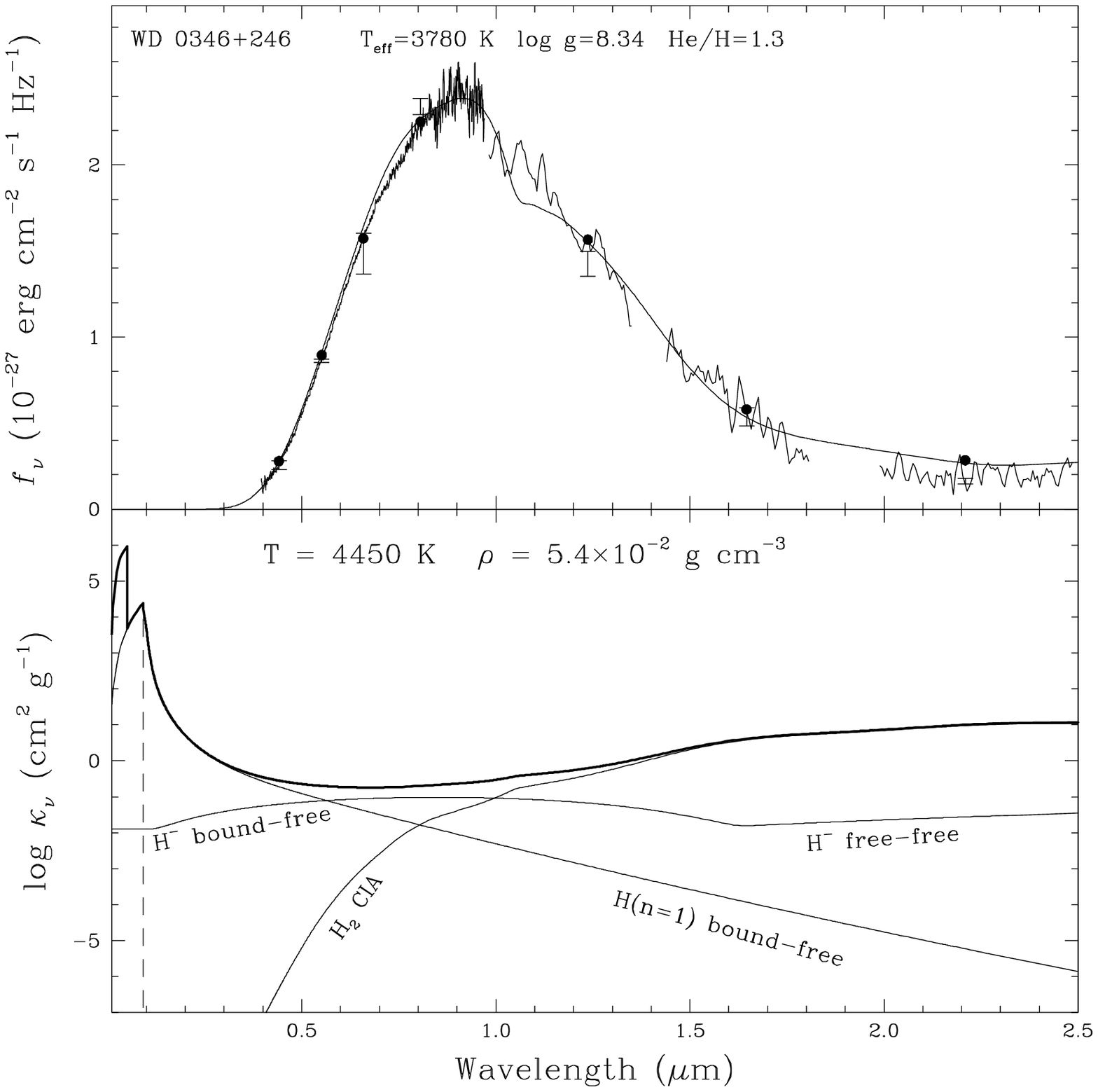] {{\it Top panel}: Our best fit to the energy 
distribution of \WD\ using a pseudo-continuum opacity calculated with
the damping function given by equation (4) with $a=0.52$ and $\Delta
\lambda_{\rm D}=36$ \AA. {\it Bottom panel}: Most important sources 
of opacity at the photosphere ($\tau_{\rm R}=1$) of the corresponding
model atmosphere. The local temperature and density are indicated in
the figure, and each opacity source is identified as well. The thick
solid line represents the total opacity. The bound-free opacity from
the $n=1$ level includes the pseudo-continuum opacity from the
dissolved levels, while the contribution from the unperturbed
bound-free absorptions alone is represented by the dashed line. The
feature near 0.05 $\mu m$ is the ionization threshold from neutral
helium.\label{fg:f6}}

\clearpage
\begin{figure}[p]
\plotone{f1.ps}
\begin{flushright}
Figure \ref{fg:f1}
\end{flushright}
\end{figure}

\begin{figure}[p]
\plotone{f2.ps}
\begin{flushright}
Figure \ref{fg:f2}
\end{flushright}
\end{figure}

\begin{figure}[p]
\plotone{f3.ps}
\begin{flushright}
Figure \ref{fg:f3}
\end{flushright}
\end{figure}

\begin{figure}[p]
\plotone{f4.ps}
\begin{flushright}
Figure \ref{fg:f4}
\end{flushright}
\end{figure}

\begin{figure}[p]
\plotone{f5.ps}
\begin{flushright}
Figure \ref{fg:f5}
\end{flushright}
\end{figure}

\begin{figure}[p]
\plotone{f6.ps}
\begin{flushright}
Figure \ref{fg:f6}
\end{flushright}
\end{figure}


\clearpage
\begin{thebibliography}{}

\bibitem[Bergeron et al.(2001)]{blr01} Bergeron, P., Leggett, S. K., \& Ruiz, M. T.
2001, \apjs, 133, 413

\bibitem[Bergeron et al.(1997)]{brl97} Bergeron, P., Ruiz, M. T., \& Leggett, S. 
K. 1997, \apjs, 108, 339 

\bibitem[Bergeron et al.(1994)]{ber94} Bergeron, P., Ruiz, M. T., Leggett, S. K., 
Saumon, D., \& Wesemael, F. 1994, \apj, 423, 456

\bibitem[Bergeron et al.(1995)]{bsw95} Bergeron, P., Saumon, D., \& Weseamel, F. 
1995, \apj, 443, 764

\bibitem[Bergeron et al.(1991)]{ber91} Bergeron, P., Weseamel, F., \& Fontaine, G.
1991, \apj, 367, 253

\bibitem[D\"appen et al.(1987)]{dappen} D\"appen, W., Anderson, L., \& Mihalas, D.
1987, \apj, 319, 195

\bibitem[Dupuis et al.(1992)]{dupuis92} Dupuis, J., Fontaine, G., Pelletier, C., 
\& Wesemael, F. 1992, \apjs, 82, 505

\bibitem[Dupuis et al.(1993)]{dupuis93} Dupuis, J., Fontaine, G., \& Wesemael, F. 
1993, \apjs, 87, 345

\bibitem[Fontaine et al.(2001)]{fon01} Fontaine, G., Brassard, P., \& Bergeron, P.
2001, \pasp, 113, 409

\bibitem[Hambly et al.(1997)]{hambly97} Hambly, N. C., Smartt, S. J., \& 
Hodgkin, S. T. 1997, \apj, 489, L157

\bibitem[Hambly et al.(1999)]{hambly99} Hambly, N. C., Smartt, S. J., Hodgkin, S. T., 
Jameson, R. F., Kemp, S. N., Rolleston, W. R. J., \& Steele I. A. 1999, \mnras, 309, 
L33

\bibitem[Hansen(1998)]{hansen98} Hansen, B. M. S. 1998, \nat, 394, 860

\bibitem[Harris et al.(1999)]{harris99} Harris, H. C., Dahn, C. C., Vrba, F. J.,
Henden, A. A., Liebert, J., Schmidt, G. D., \& Reid, I. N. 1999, \apj, 524, 1000

\bibitem[Harris et al.(2001)]{harris01} Harris, H. C., et al. 2001, \apj, 549, L109

\bibitem[Hummer \& Mihalas(1988)]{hm88} Hummer, D. G., \& Mihalas, D. 1988, 
\apj, 331, 794

\bibitem[Ibata et al.(2000)]{ibata00} Ibata, R., Irwin, M., Bienaym\'e, O., Scholz, R.,
\& Guibert, J. 2000, \apj, 532, L41

\bibitem[J\o rgensen et al.(2000)]{jorgensen} J\o rgensen, U. G., Hammer, D., Borysow,
 A., \& Falkesgaard, J. 2000, \aap, 361, 283

\bibitem[Knox et al.(1999)]{knox99} Knox, R. A., Hawkins, M. R. S., \& Hambly, N. C. 
1999, \mnras, 306, 736

\bibitem[Lenzuni et al.(1991)]{lenzuni} Lenzuni, P., Chernoff, D. F., \& Salpeter, 
E. E. 1991, \apjs, 76, 759

\bibitem[Liebert et al.(1988)]{ldm88} Liebert, J., Dahn, C. C., \& Monet, D. G. 
1988, \apj, 332, 891

\bibitem[Malo et al.(1999)]{malo99} Malo, A., Wesemael, F., \& Bergeron P. 1999, 
\apj, 517, 901

\bibitem[Monet et al.(1992)]{monet92} Monet, D. G., Dahn, C. C., Vrba, F. J., 
Harris, H. C., Pier, J. R., Luginbuhl, C. B., \& Ables, H. D. 1992, \aj, 103, 638

\bibitem[Oppenheimer et al.(2001a)]{opp01a} Oppenheimer, B. R., Hambly, N. C., 
Digby, A. P., Hodgkin, S. T., \& Saumon, D. 2001a, Science, in press

\bibitem[Oppenheimer et al.(2001b)]{opp01b} Oppenheimer, B. R., Saumon, D., 
Hodgkin, S. T., Jameson, R. F., Hambly, N. C., Chabrier, G., Filipenko, A. V.,
Coil, A. L., \& Brown, M. E. 2001b, \apj, 550, 448

\bibitem[Reid et al.(2001)]{reid01} Reid, I. N., Sahu, K. C., \& Hawley, S. L. 2001,
\apj, submitted

\bibitem[Saumon \& Jacobson(1999)]{saumon99} Saumon, D., \& Jacobson, S. B. 
1999, \apjl, 511, L107

\bibitem[Tassoul et al.(1990)]{tassoul90} Tassoul, M., Fontaine, G., \& Winget, 
D. E. 1990, \apjs, 72, 335

\bibitem[Wesemael(1979)]{wes79} Wesemael, F. 1979, \aap, 72, 104

\bibitem[Zeidler-K.~T. et al.(1986)]{zeidler86} Zeidler-K.~T., E.-M., Weidemann, 
V., \& Koester, D. 1986, \aap, 155, 356

\end{thebibliography}
\end{document}